\documentclass{DISproc}

\begin{document}
%------------------------------------
\title{Production of $c \bar c$ pairs at LHC: $k_t$-factorization and
double-parton scattering}

%for single authors the superscripts are optional
\author{{\slshape Antoni Szczurek$^1$,$^2$}\\[1ex]
$^1$Institute of Nuclear Physics PAN, ul. Radzikowskiego 152, PL-31-342 Krak\'ow, Poland\\
$^2$Rzesz\'ow University, ul. Rejtana 16, PL-35-959 Rzesz\'ow, Poland }

% please enter the contribution ID for the DOI
\contribID{xy}

\doi  % if there is an online version we will register DOIs

\maketitle

\begin{abstract}
We discuss charm production at LHC. The production of single $c \bar c$ pairs
is calculated in the $k_t$-factorization approach. We use several
unintegrated gluon distributions from the literature.
Differential distributions for several charmed mesons are presented and 
compared to recent results of the ALICE and LHCb collaborations.
Some missing strength can be observed.
Furthermore we discuss production of two $c \bar c$ pairs within a simple
formalism of double-parton scattering (DPS). Surprisingly large
cross sections, comparable to single-parton scattering (SPS)
contribution to $c \bar c$ production, are predicted for LHC energies. 
\end{abstract}

%----------------------------
\section{Introduction}
%----------------------------

The cross section for open charm production at the LHC is very large.
Different mesons are measured \cite{ALICE_charm,LHCb_charm}. Some other 
experiments are preparing their experimental cross sections. 
Different theoretical approaches for heavy quark production are
used in the literature. In the present communication we present
briefly some results for charmed meson production within
$k_t$-factorization approach. A more detailed analysis will be presented
elsewhere \cite{MS2012}. Similar analysis within next-to-leading order
aproach was presented very recently \cite{KKSS2012}.
Previously we have used the $k_t$-factorization approach for charm
production at the Tevatron \cite{LS2006} and for nonphotonic electron 
production at RHIC \cite{LMS2009,MSS2011}.
The $k_t$-factorization approach was also successfully used for beauty
\cite{JKLZ2011} and top \cite{LZ2011} quark (antiquark) inclusive production.

Recently we have made first estimates for the production of two $c \bar
c$ pairs \cite{LMS2011,SS2012}. We have considered both double-parton 
scattering (DPS) mechanism \cite{LMS2011} as well as single-parton scattering 
(SPS) mechanism \cite{SS2012}. By comparison of contributions of both 
mechanisms
we come to the conclusion that the production of two $c \bar c$ pairs
is a favourite place to study and identify double-parton scattering effects.
The double-parton scattering was studied recently for different
high-energy processes.

%-------------------------------------------------------------
\section{Inclusive charmed meson production}
%-------------------------------------------------------------

In the leading-order (LO) approximation within the $k_t$-factorization approach
the quadruply differential cross section in the rapidity 
of $Q$ ($y_1$), in the rapidity of $\bar Q$ ($y_2$) and in the transverse 
momentum of $Q$ ($p_{1,t}$) and $\bar Q$ ($p_{2,t}$) can be written as
\begin{eqnarray}
\frac{d \sigma}{d y_1 d y_2 d^2p_{1,t} d^2p_{2,t}} =
\sum_{i,j} \; \int \frac{d^2 \kappa_{1,t}}{\pi} \frac{d^2 \kappa_{2,t}}{\pi}
\frac{1}{16 \pi^2 (x_1 x_2 s)^2} \; \overline{ | {\cal M}_{ij \to Q \bar Q} |^2}\\
\nonumber 
\delta^{2} \left( \vec{\kappa}_{1,t} + \vec{\kappa}_{2,t} 
                 - \vec{p}_{1,t} - \vec{p}_{2,t} \right) \;
{\cal F}_i(x_1,\kappa_{1,t}^2) \; {\cal F}_j(x_2,\kappa_{2,t}^2) \; , 
\nonumber  
\end{eqnarray}
where ${\cal F}_i(x_1,\kappa_{1,t}^2)$ and ${\cal F}_j(x_2,\kappa_{2,t}^2)$
are so-called unintegrated gluon (parton) distributions. 
The unintegrated parton distributions are evaluated at:
$x_1 = \frac{m_{1,t}}{\sqrt{s}}\exp( y_1) 
     + \frac{m_{2,t}}{\sqrt{s}}\exp( y_2)$,
$x_2 = \frac{m_{1,t}}{\sqrt{s}}\exp(-y_1) 
     + \frac{m_{2,t}}{\sqrt{s}}\exp(-y_2)$,
where $m_{i,t} = \sqrt{p_{i,t}^2 + m_Q^2}$.

\begin{figure}[!h]
\includegraphics[width=6cm]{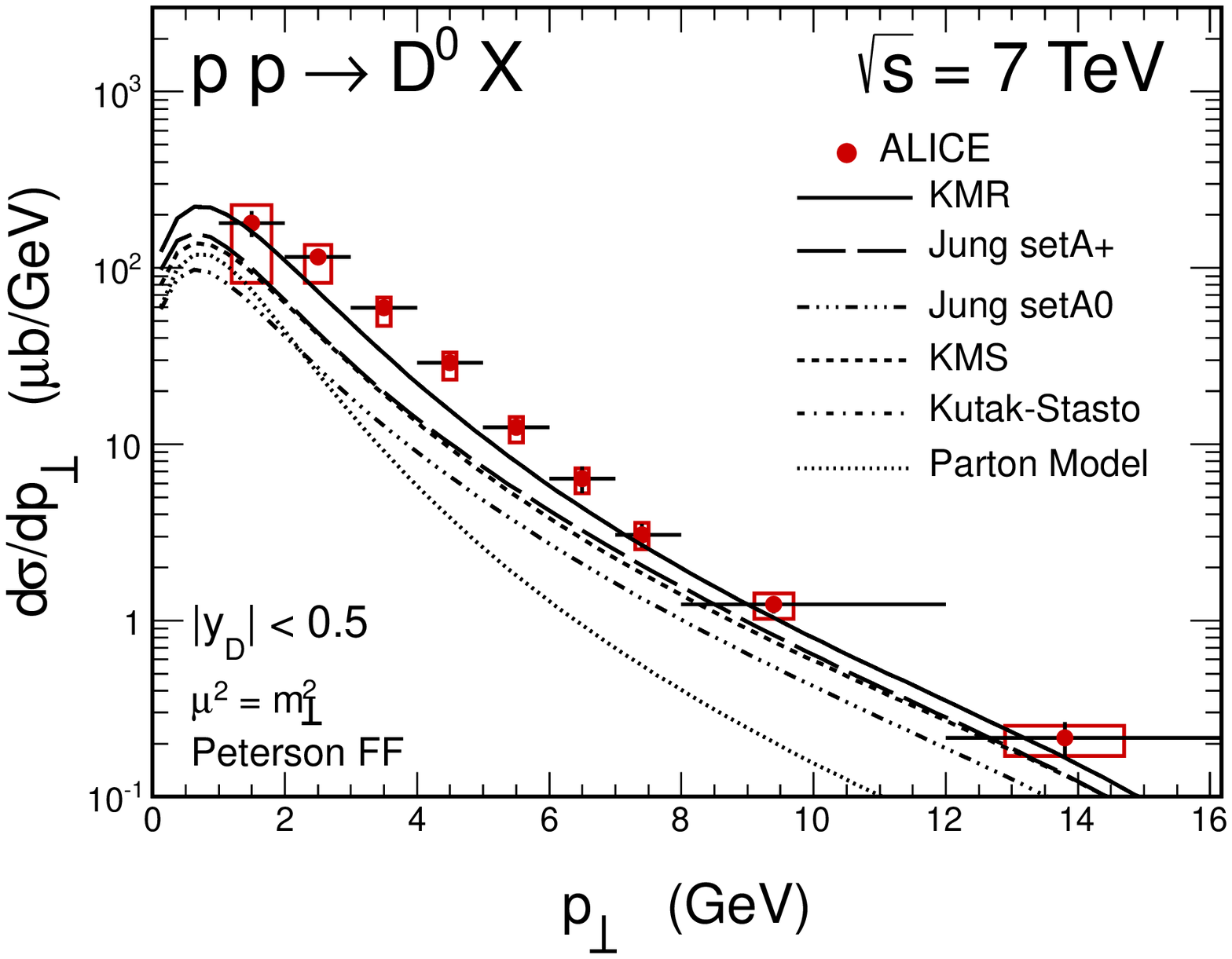}
\includegraphics[width=6cm]{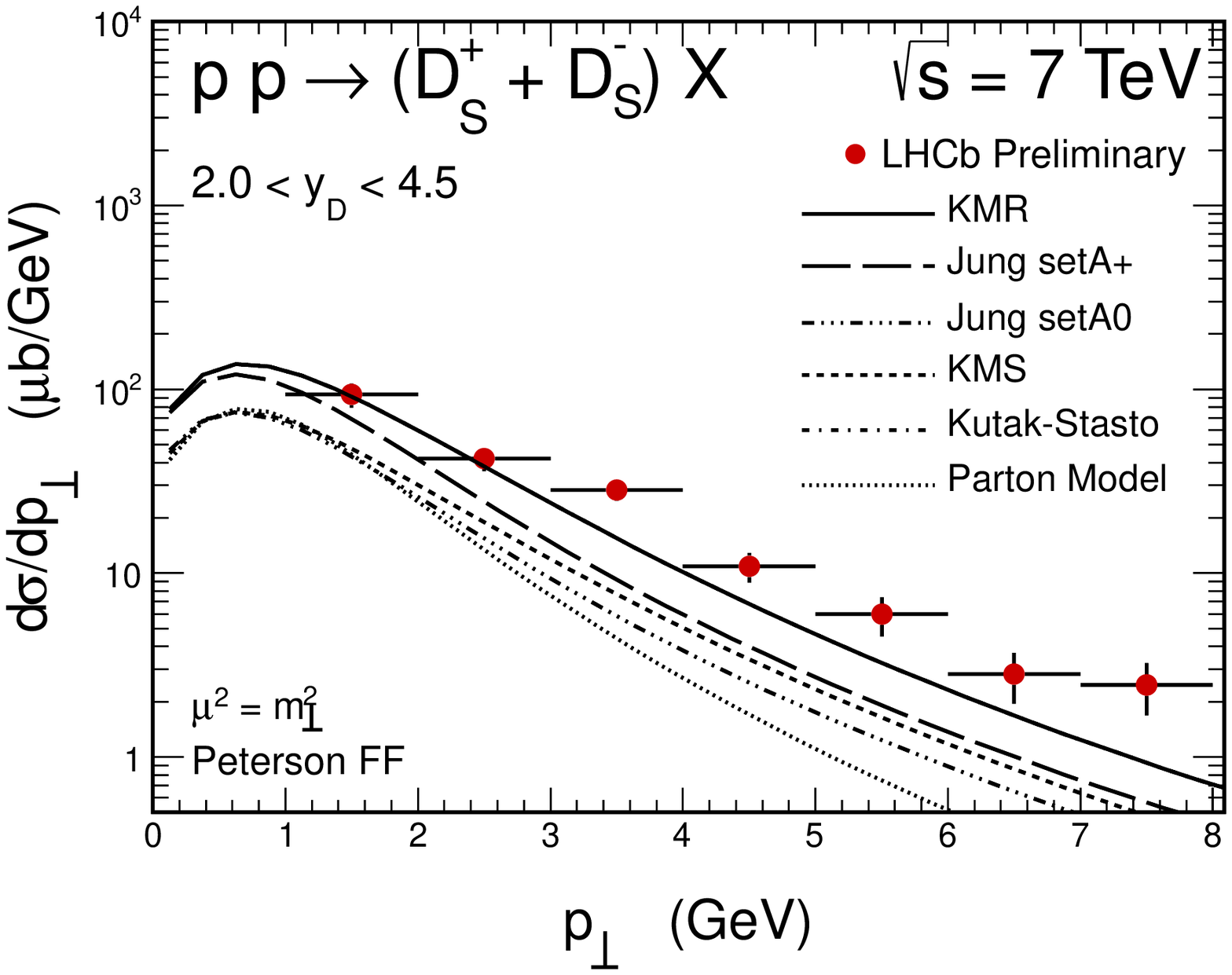}
   \caption{
\small Two examples of transverse momentum distribution of charmed
mesons compared to ALICE (left panel) and LHCb (right panel)
experimental data. The calculations were done for different unintegrated
gluon distributions.
}
 \label{fig:dsig_dpt_mesons}
\end{figure}

The hadronization is done in the way explained in Ref.\cite{LMS2009}.
In Fig.\ref{fig:dsig_dpt_mesons} we show two examples how we describe LHC
experimental data \cite{ALICE_charm,LHCb_charm}. There seems to be some 
strength missing. A possible explanation of that observation will be 
discussed in the next section.

%-----------------------------------------------------------------------------
%------------------------------------------------------------------------------

%---------------------------------------------------------
\section{Production of two $c \bar c$ pairs}
%---------------------------------------------------------

Two possible mechanisms of the production of two $c \bar c$ pairs are shown in
Fig.\ref{fig:mechansims_ccbarccbar}.
%-----------------------------------------------------------------------------
\begin{figure}[!h]
\begin{center}
\includegraphics[width=4cm]{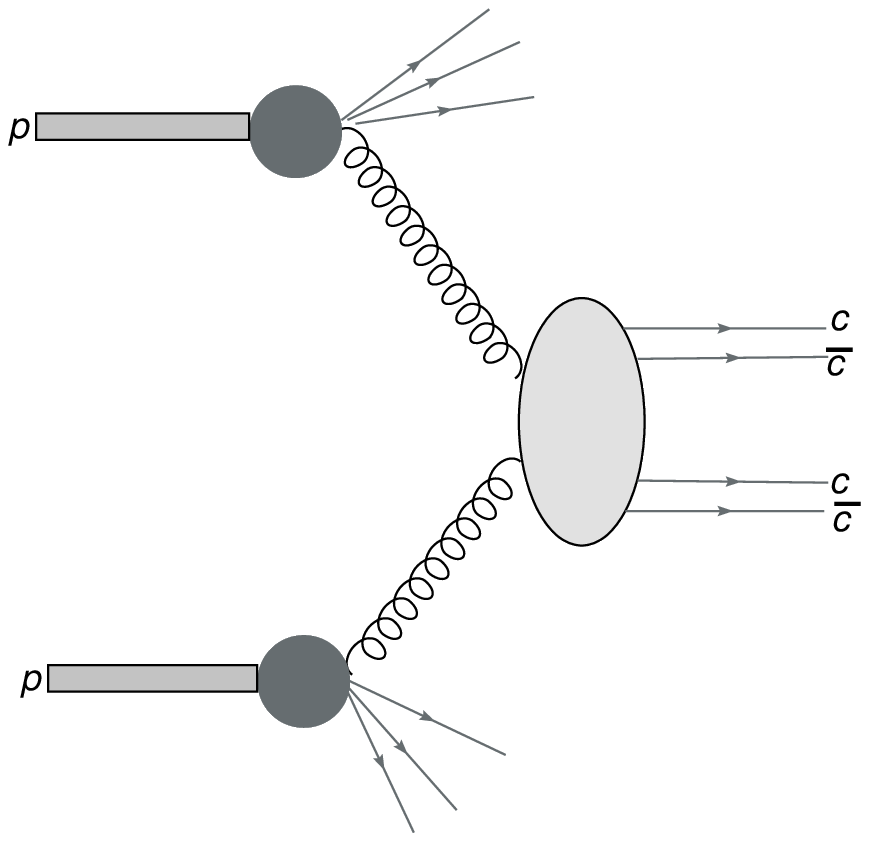}
\includegraphics[width=4cm]{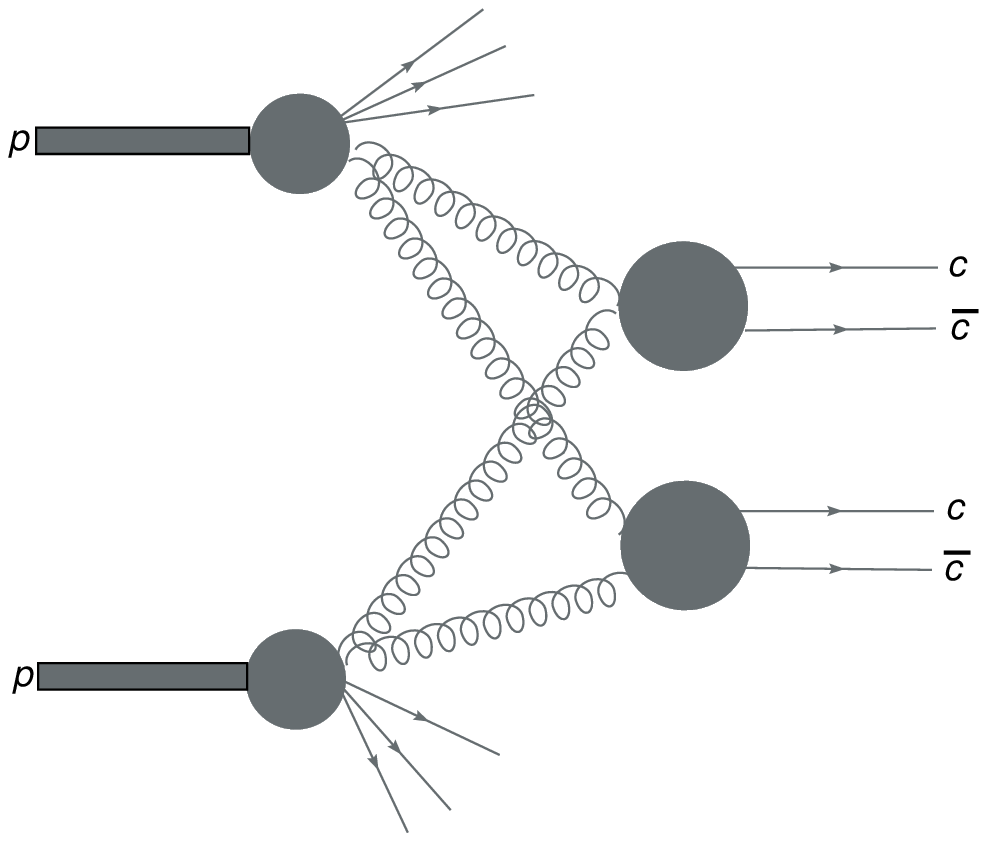}
\end{center}
   \caption{
\small SPS (left) and DPS (right) mechanisms of $(c \bar c) (c \bar c)$ 
production.  
}
\label{fig:mechanisms_ccbarccbar}
\end{figure}
%------------------------------------------------------------------------------

The cross section for differential distribution in a simple
double-parton scattering in leading-order collinear approximation 
can be written as
\begin{equation}
\frac{d \sigma}{d y_1 d y_2 d^2 p_{1t} d y_3 d y_4 d^2 p_{2t}}  \\ =
\frac{1}{ 2 \sigma_{eff} }
\frac{ d \sigma } {d y_1 d y_2 d^2 p_{1t}} \cdot
\frac{ d \sigma } {d y_3 d y_4 d^2 p_{2t}} 
\label{differential_distribution}
\end{equation}
which by construction reproduces the formula for integrated cross
section \cite{LMS2011}.
This cross section is formally differential in 8 dimensions but can be 
easily reduced to 7 dimensions noting that physics of unpolarized
scattering cannot depend on azimuthal angle of the pair or on azimuthal
angle of one of the produced $c$ ($\bar c$) quark (antiquark).
This can be easily generalized by including QCD evolution effects
\cite{LMS2011}.

\begin{figure}[!h]
\begin{center}
\includegraphics[width=6.0cm]{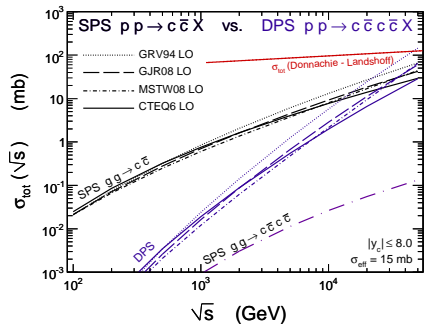}
\end{center}
   \caption{
\small Total LO cross section for single
$c \bar c$ pair and SPS and DPS $c \bar c c \bar c$ production
as a function of center-of-mass energy.  
}
 \label{fig:single_vs_double_LO}
\end{figure}

In Fig.~\ref{fig:single_vs_double_LO} we
compare cross sections for the single $c \bar c$ pair production as well
as for single-parton and double-parton scattering $c \bar c c \bar c$
production as a function of proton-proton center-of-mass energy. 
At low energies the conventional single $c \bar c$ pair production
cross section is much larger. 
The cross section for SPS production
of $c \bar c c \bar c$ system is more than two orders of magnitude smaller
than that for single $c \bar c$ production. For reference we show the
proton-proton total cross section as a function of energy.
At higher energies the DPS contribution
of $c \bar c c \bar c$ quickly approaches that for single $c \bar c$ 
production as well as the total cross section.

%-----------------------------------------------------------------------------
%-----------------------------------------------------------------------------

\begin{figure}[!h]
\begin{center}
\includegraphics[width=5.0cm]{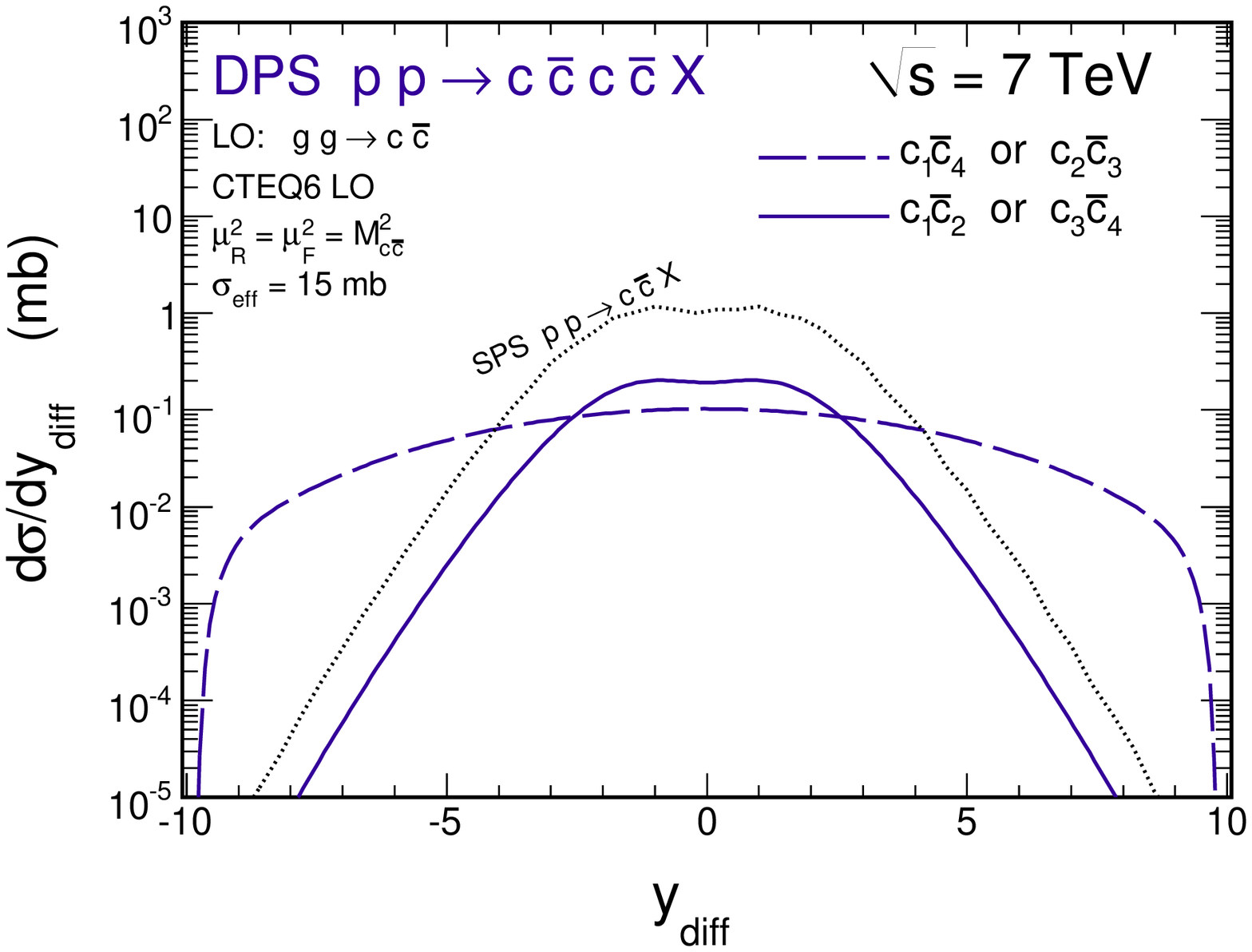}
\includegraphics[width=5.0cm]{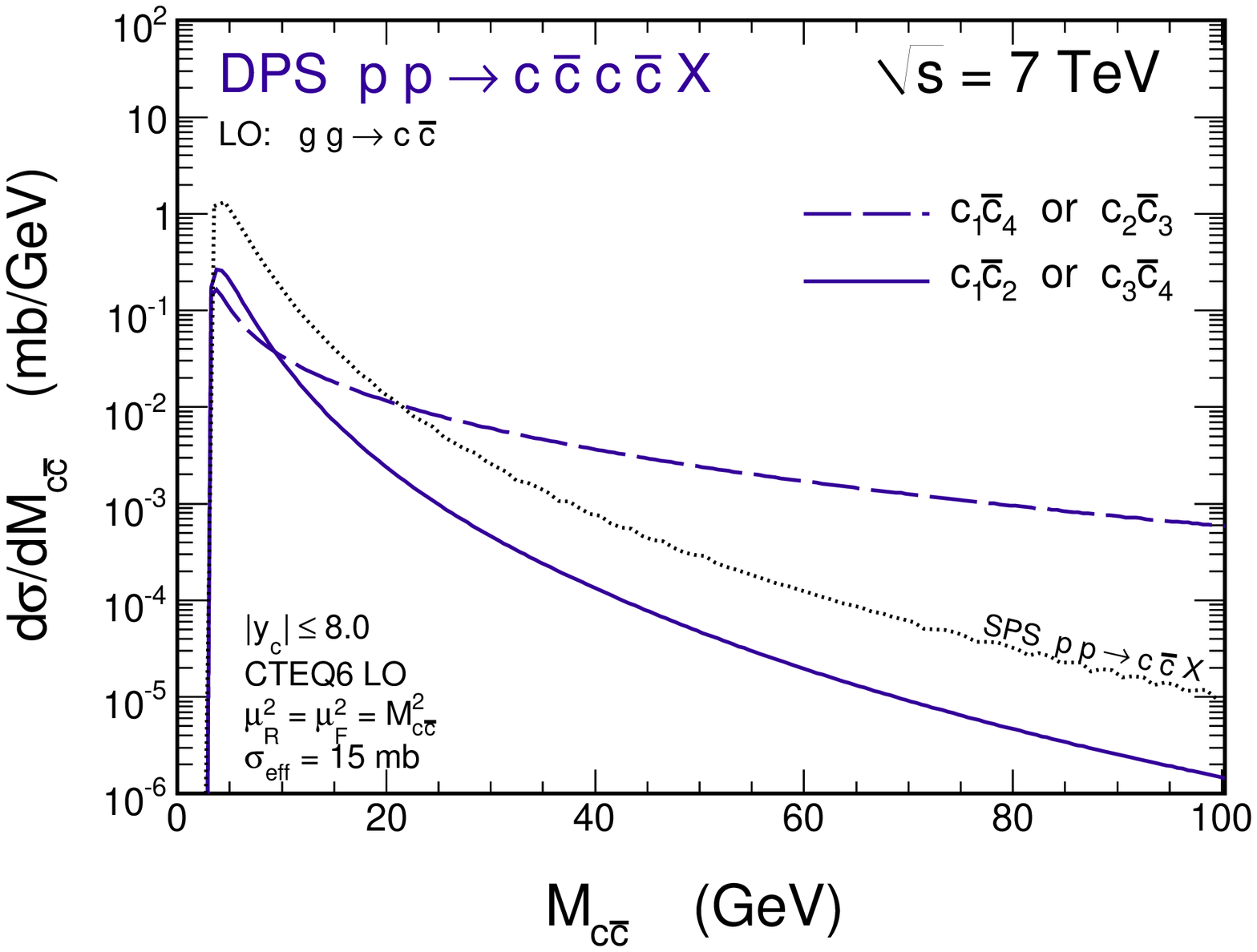}
\end{center}
   \caption{
\small Distribution in rapidity difference (left panel) and in invariant
mass of the $c\bar{c}$ pair (right panel) at $\sqrt{s}$ = 7 TeV.
}
\label{fig:double_correlations}
\end{figure}

In Ref.\cite{LMS2011} we have also considered several correlation
observables between different $c$ quarks and $\bar c$ antiquarks.
Particularly interesting are correlations between $c$-$c$ and $\bar
c$-$\bar c$. Two examples are shown in Fig.\ref{fig:double_correlations}.
We show both terms: when 
$c \bar c$ are emitted in the same parton scattering 
($c_1\bar c_2$ or $c_3\bar c_4$) and when they are emitted from different 
parton scatterings ($c_1\bar c_4$ or $c_2\bar c_3$). In the latter case
we observe a long tail for large rapidity difference as well as at large
invariant masses of $c \bar c$.

%-----------------------------------------------------------------------------
%------------------------------------------------------------------------------

\begin{figure}[!h]
\begin{center}
\includegraphics[width=5.0cm]{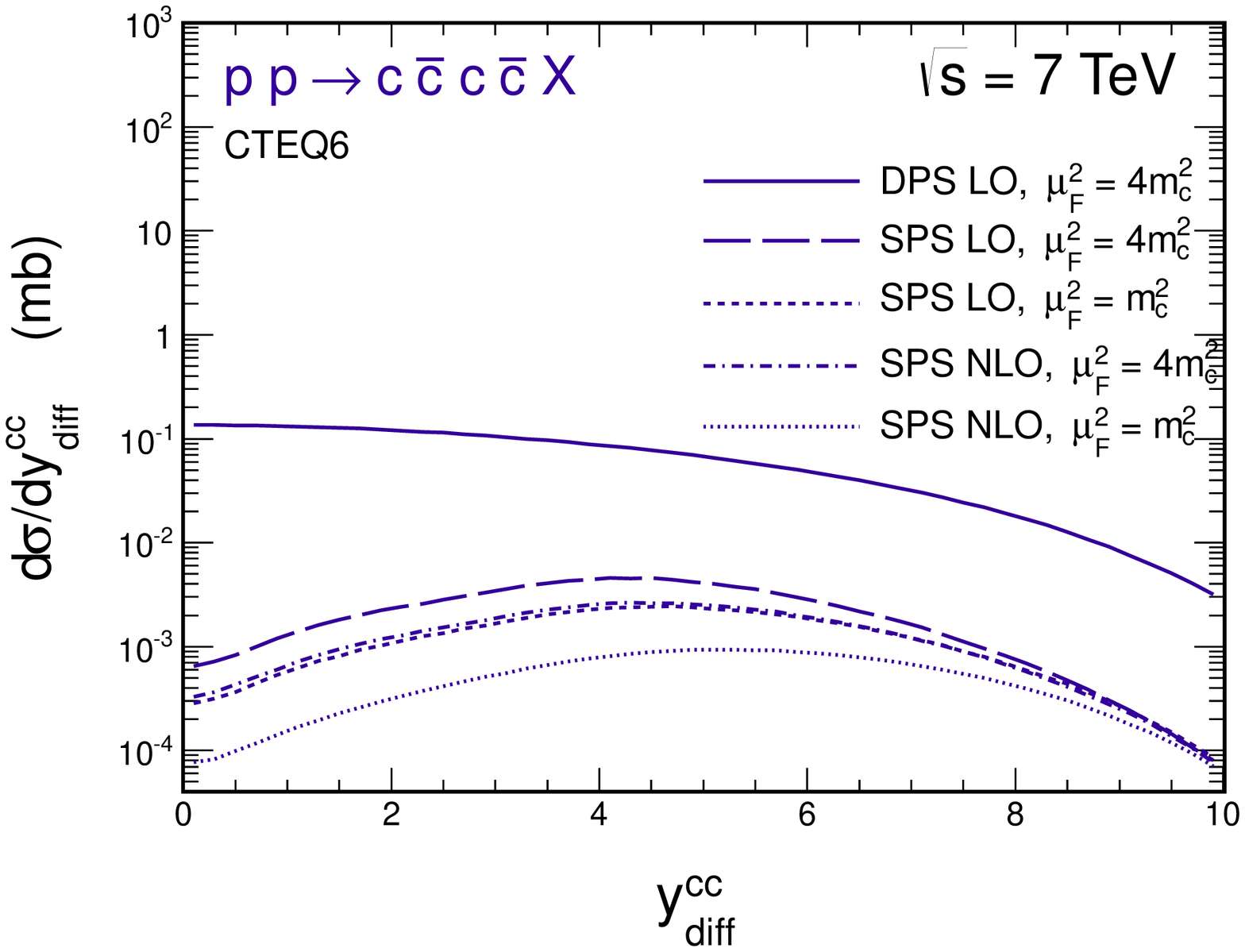}
\includegraphics[width=5.0cm]{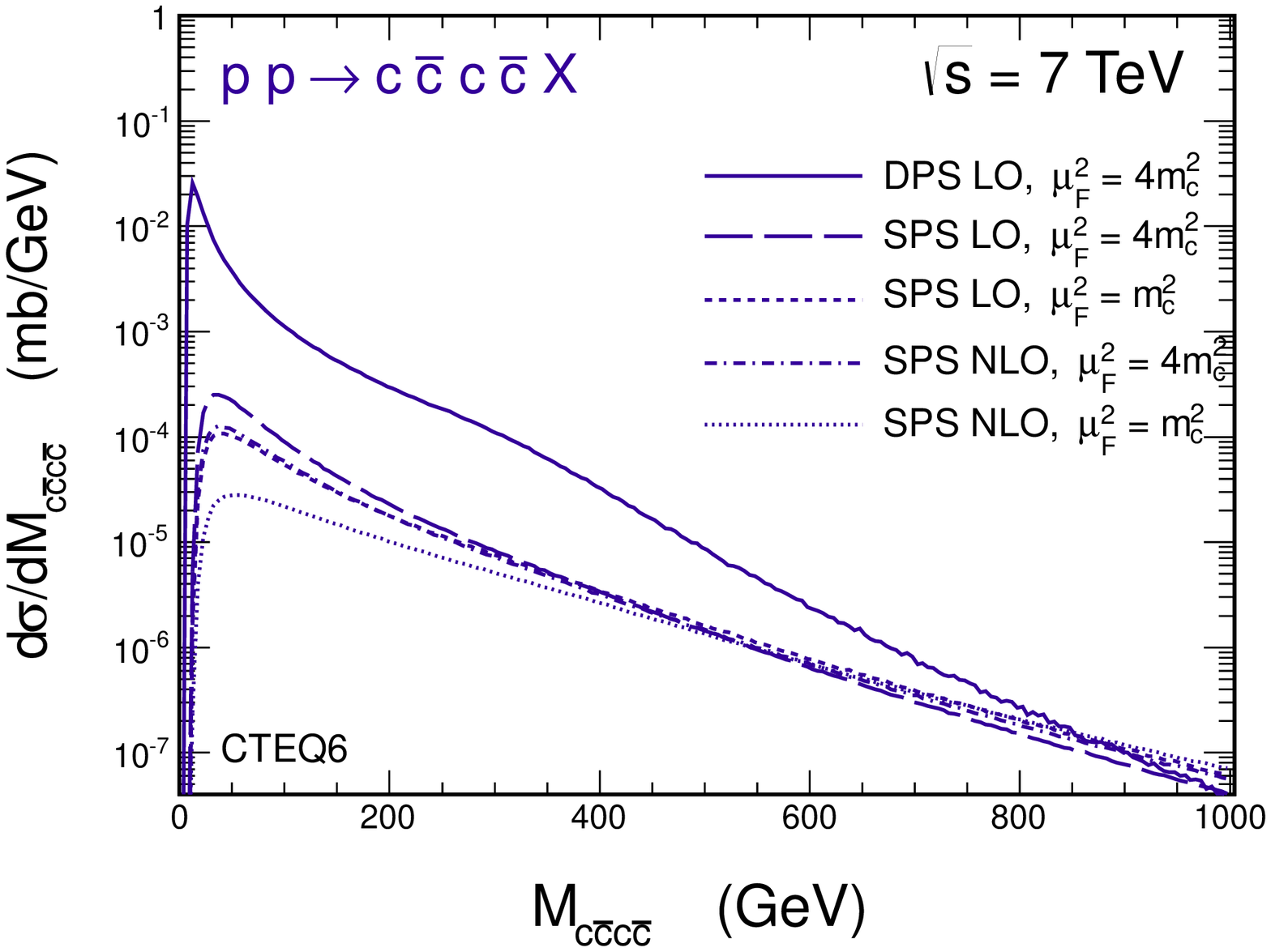}
\end{center}
   \caption{Comparison of SPS and DPS contributions for two
correlation distributions.
}
 \label{fig:correlation_SPS_vs_DPS}
\end{figure}

In Ref.\cite{SS2012} we have calculated cross section for 
$c \bar c c \bar c$ production in single-parton scattering in
high-energy approximation. In Fig.\ref{fig:correlation_SPS_vs_DPS}
we compare the SPS contribution with the DPS one. Clearly the SPS
contribution at large rapidity difference between $c c$ or $\bar c \bar
c$ is much smaller than the DPS contribution.

%-----------------------------------------------------------------------------
%------------------------------------------------------------------------------

\begin{figure}[!h]
\begin{center}
\includegraphics[width=5.0cm]{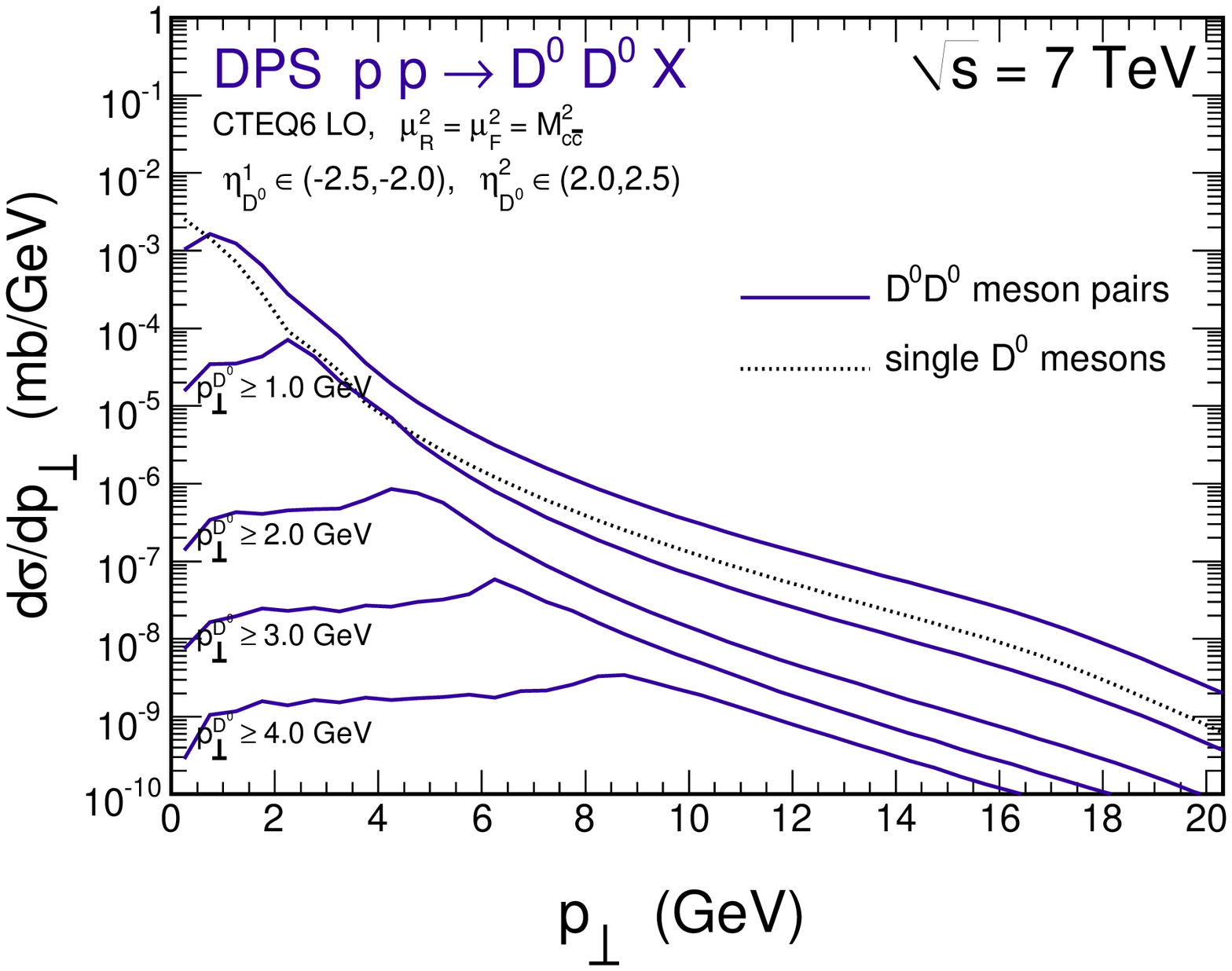}
\includegraphics[width=5.0cm]{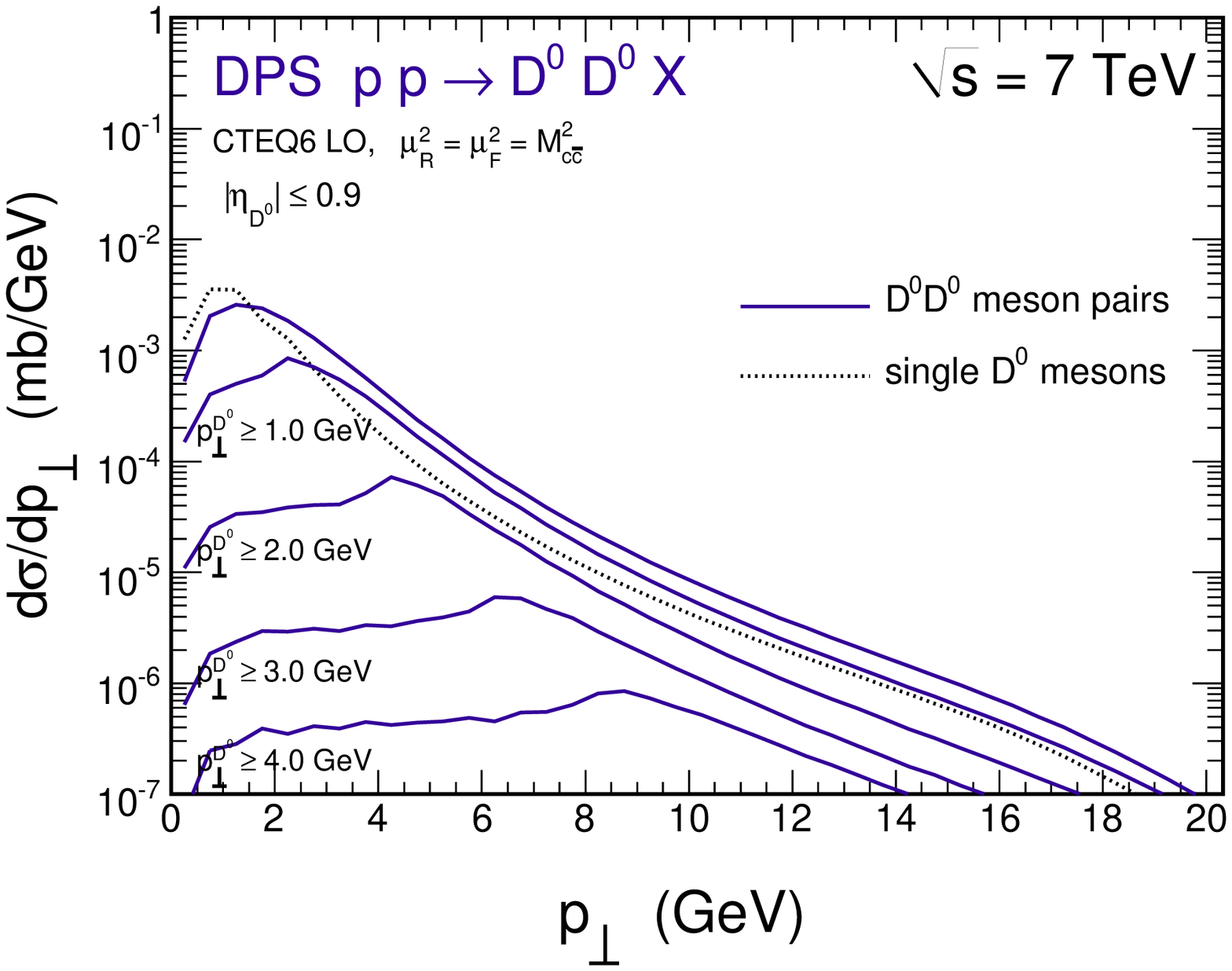}
\end{center}
   \caption{
\small Transverse momentum distribution of the $D^0 D^0$ (or ${\bar D}^0
{\bar D}^0$) pairs for the rapidity interval of 
ATLAS or CMS (left) and ALICE (right) experiments for different cuts 
on transverse momenta of each meson in the pair.
The distribution in transverse momentum of single $D^0$ is shown for 
comparison (dashed line). 
All distributions are shown for $\sqrt{s}$ = 7 TeV. 
}
 \label{fig:dsig_dptsum_mesons}
\end{figure}

In Ref.\cite{LMS2011} we have discussed that the sum of transverse
momenta of two $c$ 
(or two $\bar c$) has a hard tail. This is of course not an observable.
In Fig.\ref{fig:dsig_dptsum_mesons} we show instead distribution in 
transverse momentum of the $D^0 D^0$ pair (or ${\bar D}^0 {\bar D}^0$
pair) for the rapidity interval relevant for a given experiment. 
This distribution has surprisingly long tail.
For comparison we show transverse momentum distribution of one $D^0$ 
(or ${\bar D}^0$).

%-----------------------------------------------------------------------------
%------------------------------------------------------------------------------

%--------------------
\section{Summary}
%--------------------

We have presented our selected new results for charmed meson production
at LHC. Results of our calculation have been compared with recent
ALICE and LHCb experimental data for transverse momentum distribution
of $D$ mesons. There seems to be a missing strength, especially
for the LHCb kinematics.

One of possible explanation is a presence of double-parton scattering 
contributions.
Therefore the second topic discussed during the conference was 
the production of two $c \bar c$ pairs.
We have compared energy dependence of the DPS contribution to 
the $c \bar c c \bar c$ production with that for the $c \bar c$
production. The cross section for two pair production grows much faster
than that for single pair production. At high energies the two cross
sections become comparable. We have also discussed some correlation
observables that could be used to identify double-parton scattering
contribution. The rapidity difference is one of the good examples.

We have also estimated corresponding single-parton scattering
contributions in a high energy approach. The latter turned out
to be much smaller than the double-parton scattering contributions.

In summary, we have found that the production of two $c \bar c$ pairs
is one of the best places to study and identify double-parton scattering
effects.
For example a good possibility would be to measure two
$D^0 D^0$ or two $\bar D^0 \bar D^0$ mesons. The LHCb collaboration
has started already such studies \cite{LHCb_DPS}.

%----------------------------------
\section{Acknowledgements}
%----------------------------------

I am indebted to Marta {\L}uszczak, Rafa{\l} Maciu{\l}a and Wolfgang
Sch\"afer for collaboration on the issues presented here.
I wish to thank the organizers for perfect organization of DIS2012. 
My stay during the conference in Bonn was supported from a polish grant
DEC-2011/01/B/ST2/04535.

%To acknowledge funding bodies etc., a special section may be placed
%before the bibliography: \verb?\section*{Acknowledgements}?.

%\section{Bibliography}

% ****************************************************************************
% BIBLIOGRAPHY AREA
% ****************************************************************************

{\raggedright
\begin{footnotesize}
% IF YOU DO NOT USE BIBTEX, USE THE FOLLOWING SAMPLE SCHEME FOR THE REFERENCES
% ----------------------------------------------------------------------------

% ----------------------------------------------------------------------------

% IF YOU USE BIBTEX,
% - DELETE THE TEXT BETWEEN THE TWO ABOVE DASHED LINES
% - UNCOMMENT THE NEXT TWO LINES AND REPLACE 'smith_joe.bib' WITH YOUR
%   FILE(S)

% \bibliographystyle{DISproc}
% \bibliography{smith_joe.bib}
\end{footnotesize}
}

% ****************************************************************************
% END OF BIBLIOGRAPHY AREA
% ****************************************************************************

\end{document}